\documentclass{PoS}

\usepackage{siunitx}
\usepackage{amsmath}
\usepackage{amssymb}

\title{The Mirror Alignment and Control System for CT5 of the H.E.S.S. experiment}

\ShortTitle{MACS for CT5 of H.E.S.S.}

\author{\speaker{Daniel Gottschall}$^a$, Andreas F\"orster$^b$, Antonio Bonardi$^c$, Andrea Santangelo$^a$, and Gerd P\"uhlhofer$^a$ for the H.E.S.S. collaboration\\
E-mail:  \email{daniel.gottschall@astro.uni-tuebingen.de}\\
\llap{$^a$} Institut f\"ur Astronomie und Astrophysik, Abteilung Hochenergieastrophysik, Kepler Center for
Astro and Particle Physics, Eberhard-Karls-Universit\"at\\
Sand 1, D 72076 T\"ubingen, Germany\\
\llap{$^b$} Max-Planck-Institut f\"ur Kernphysik\\
P.O. Box 103980, D 69029 Heidelberg, Germany\\
\llap{$^c$} Department of Astrophysics, Radboud University\\
Heyendaalseweg 135, 6525 AJ Nijmegen, Netherlands\\

}

\abstract{The High Energy Stereoscopic System (H.E.S.S.) experiment is one of the largest observatories for gamma-ray astronomy. It consists of four telescopes with a reflecting dish diameter of \SI{12}{\metre} (CT1 to CT4) and a newer large telescope (CT5) with a reflecting dish diameter of \SI{28}{\metre}. On CT5 876 mirror facets are mounted, all of them equipped with a computerised system for their alignment. The design of the mirror alignment and control system and the performance of the hardware installed to the telescope are presented. Furthermore the achieved point spread function of the telescope over the full operational elevation range as well as the stability of the alignment over an extended period of time are shown.}

\FullConference{The 34th International Cosmic Ray Conference,\\
		30 July- 6 August, 2015\\
		The Hague, The Netherlands}

\begin{document}

\section{Introduction}
H.E.S.S. is a stereoscopic system of five imaging atmospheric Cherenkov telescopes in the Khomas Highlands of Namibia. The experiment consists of four telescopes with a reflecting dish diameter of \SI{12}{\metre} (CT1 to CT4) and a newer large telescope (CT5) with a reflecting dish diameter of \SI{28}{\metre}. This proceeding deals with the mirror alignment and control system of CT5. The performance of the gamma-ray observations of H.E.S.S. highly depends on a precise alignment of all 876 mirror facets of the telescope.  The large number or mirrors requires a robust system with a large degree of automation, a high precision and stable results over a long period to avoid the need of regular realignments. Since the dish of the telescope is sufficiently stable an active alignment during observationis not necessary. Using the experience with CT1-CT4 the mirror alignment and control system described by Cornils et al. (2003) \cite{Cornils} has been adapted and further developed for CT5 by the Institute for Astronomie and Astrophysics T\"ubingen (electronics and software) and the Max Plack Institute for Nuclear Physics in Heidelberg (mechanical parts). 

\section{Mechanics and electronics}
\begin{figure}
\begin{center}
     \includegraphics[width=.6\textwidth]{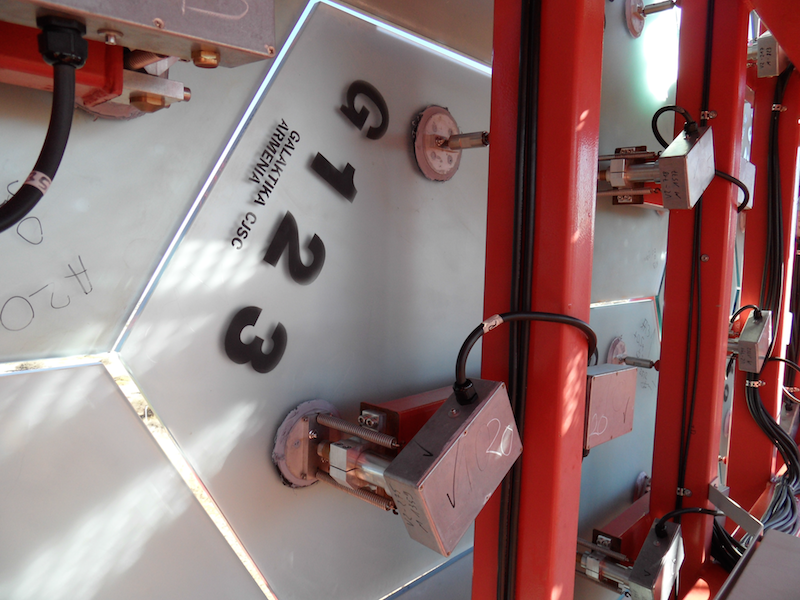}
 \end{center}
    \caption{Mechanical connection between a mirror and the dish structure: In this picture the backside of a mirror is shown. On top of the mirror the fixed connection with two rotatable ball joints is visible. On the bottom of the mirror, on the left hand side an actuator with its motor-box screwed to the back plate of the mirror is shown.}
     \label{fig1}
     \end{figure}

Every mirror is connected to the dish at three points, two of them equipped with an actuator to tilt the mirrors and one tilting but otherwise passive connector such that no stress is induced in the mirrors. The mirror connection points consist of aluminium plates glued to the backside of the mirrors. To protect the glue from sunlight a UV-resistant paste was applied. These plates are screwed to two actuators and to the tilting connector that consists of two ball joints (Figure \ref{fig1}). The spindle of the actuators has a pitch of \SI{1}{\milli\metre} and a range of \SI{40}{\milli\metre}. A DC motor including a hall sensor to count turns of the motor are connected to the spindle using a worm gear with a gear ratio of 210/1. Since both the falling and rising edge of the hall signal are detected, one turn of the spindle is separated into \SI{420}{steps}. By that the lift of the actuator can be controlled to \SI{2.4}{\micro\metre}. This translates to an angular resolution of \SI{2.2}{arcsec} which corresponds to $\frac{1}{100}$ of the pixel size of the Cherenkov camera. Highly accelerated life-time tests in the laboratory have shown that the hall signal counting system is not missing more than \SI[separate-uncertainty = true]{4(5)}{counts} over the expected lifet-time of an actuator. Further tests have shown that the difference between the actual actuator movement and the commanded movement is  \SI[separate-uncertainty = true]{0(2)}{counts}.\\
The mirrors are organised in 25 panels consisting of up to 42 mirrors per panel. Each panel hosts one panel control box equipped with control electronics. The electronics to control the actuators follows a hierarchical design starting with a motor control board at the bottom. The motor board consists of connectors for 12 motors, 12 motor drivers, a multiplexer and demultiplexer and a connector to the panel control board. The motor drivers provide a way to selected the direction of the motor movement and to brake the motor. The motor board is connected to the panel control board by one power line that is distributed to the active motor by the demultiplexer and five signals, four to select the hall-sensor and one for the hall-sensor output. This is handled by the multiplexer. One motor board can control up to 6 mirrors, for one panel up to 7 boards are needed \cite{Schwarzburg}.\\
    \begin{figure}
\begin{center}
     \includegraphics[width=.8\textwidth]{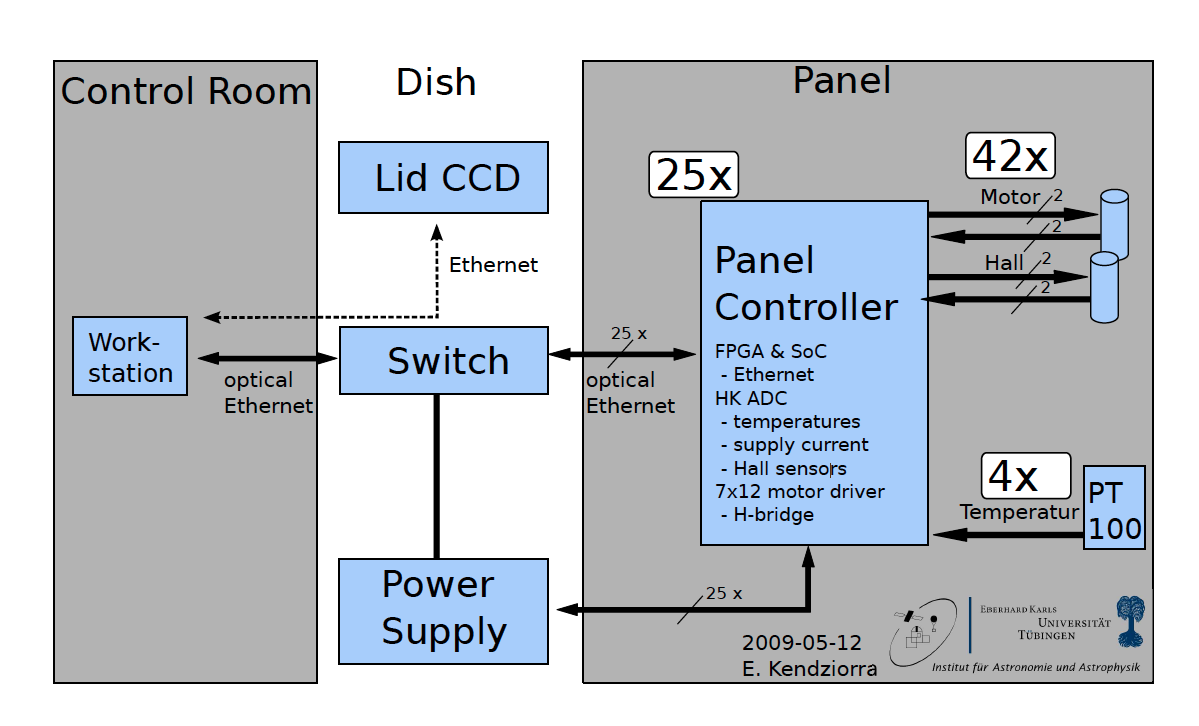}
 \end{center}
    \caption{Layout of the electronics for the mirror alignment: Every motor is connected to a motor control board with a power line and a signal line for the hall-sensor.The motor boards are part of the panel control box which also includes the panel control board equipped with an FPGA to control the motor boards. The FPGA also includes an Ethernet controller to communicate via optical fibres with the workstation in the control building. To distribute the Ethernet cables and to supply the motors with power a central control box with a switch and a power supply is located at the center of the telescope.}
     \label{fig2}
     \end{figure}
Each panel has its own panel control box consisting of the motor control boards plus one panel control board. The panel control board is equipped with a Suzaku board from Atmark Techno hosting a Spartan-3 field programming grid array (FPGA) from Xilinx. In the FPGA a micro controller with an interface to the motor control boards, with an Ethernet interface, with an interface to the analogue to digital converter (ADC) and with a fast memory to store the number of hall counts is implemented.\\
The 25 panel control boxes are located in the dish and communicate via an optical Ethernet system with the central control box where a connection to the network of the HESS experiment is established. From the central control box the power is distributed to the panel control boxes by a network controlled power supply. A schematic of this design is shown in Figure \ref{fig2} and described in detail by Schwarzburg (2012) \cite{Schwarzburg2}. The hardware to control the motors and to read out the hall counts is designed in a way that 25 mirrors can be moved at the same time.

\section{Alignment procedure}
\begin{figure}
\begin{center}
     \includegraphics[width=.6\textwidth]{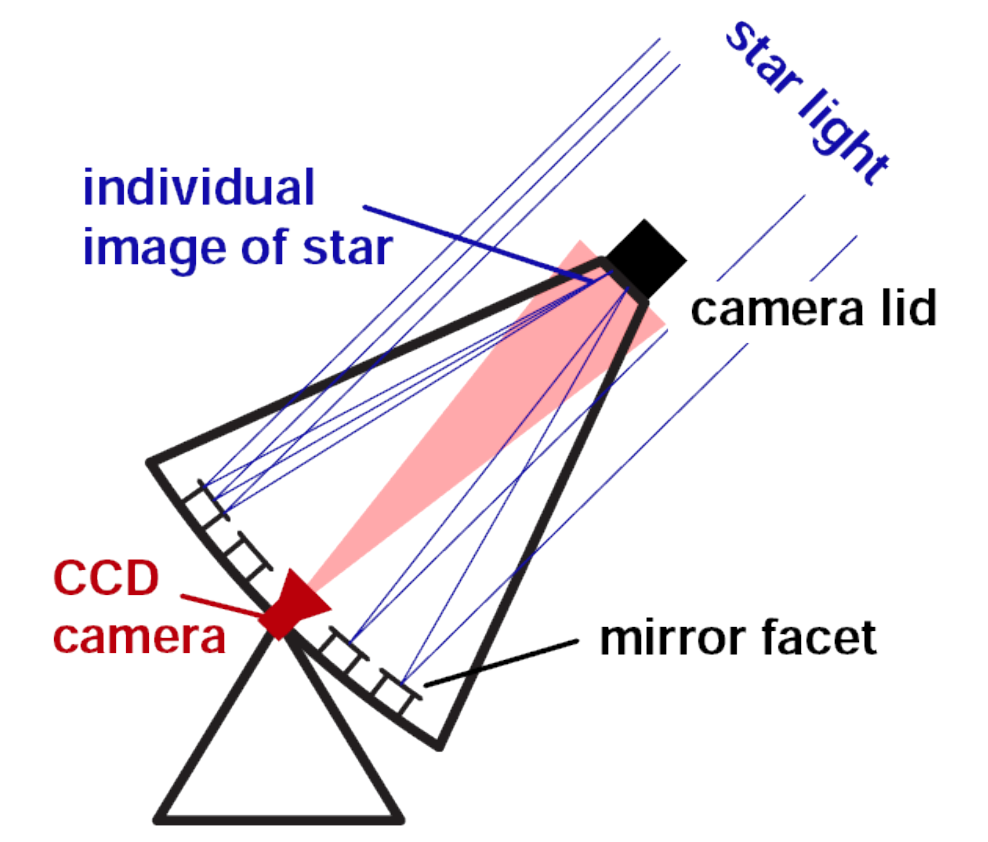}
 \end{center}
    \caption{General setup of the mirror alignment: The star light (blue) is reflected by the mirror facets to the camera lid as individual images of the star. The mirror facets are equipped with two motorised actuators that are controlled by the mirror alignment software. The CCD camera (red) in the center of the dish is taking images of the lid to identify the star images for the alignment and to measure the size of the image of the star after the alignment for monitoring.}
     \label{fig3}
     \end{figure}

For the mirror alignment a bright and isolated star is tracked by the telescope, and its image on the lid of the Cherenkov camera, viewed by a CCD camera located in the center of the dish, is used. The light of the star is reflected by every single mirror onto the lid producing individual light-spots. The goal is to align all mirrors such that they reflect the light onto the same spot at the center of the lid of the Cherenkov camera, hence defining the optical axis of the telescope \cite{Schwarzburg}. Figure \ref{fig3} is showing a diagram of the general setup of the mirror alignment.\\
For the alignment an image of the lid is taken before one or more mirrors are moved, after that another image is taken. A spot finding algorithm is implemented in software (Python-based) comparing the two images by subtracting them and searching for a significant negative and positive excess to identify the prior and the current spot position in the CCD image. From this CCD positions and the actuator positions a transformation matrix is calculated. Here, $x_1, x_2$ is defined as the position of a mirror spot on the lid and $a_1, a_2$ are the positions of the actuator of the corresponding mirror.

\begin{align} 
\Delta \vec{x}&=T\Delta \vec{a} \\
T&=\begin{pmatrix}
\delta x_1/\delta a_1 & \delta x_1/\delta a_2 \\
\delta x_2/\delta a_1 & \delta x_2/\delta a_2 
\end{pmatrix}
 \end{align}

The alignment procedure is divided into two steps, a coarse and a fine alignment. For the coarse alignment a fixed angle of \SI{120}{\degree} between the two tilting axis is assumed so that only two mirror positions are necessary to measure the full transformation matrix.

\begin{align} 
\begin{pmatrix}
\delta x_1/\delta a_1  \\
\delta x_2/\delta a_1 
\end{pmatrix}
&\approx R_{120}
\begin{pmatrix}
 \delta x_1/\delta a_2 \\
 \delta x_2/\delta a_2 
\end{pmatrix}
 \end{align}

$R_{120}$ stands for the rotation matrix for \SI{120}{\degree}. A fast alignment algorithm has been implemented that permits to coarse align up to 25 mirrors at a time. After the transformation matrix for each mirror has been found, the corrosponding mirror is moved such that the image of the star is no longer in the field of view of the CCD camera. During the fine alignment each mirror is moved to four positions around the alignment position with an offset of 200 hall count which corrosponds to \SI{440}{arcsec}. A fine alignment transformation matrix is calculated and the mirror spot is moved either to the center of the Cherenkov camera or to the center of gravity of an already existing main spot. This procedure can be done with up to four mirrors at a time.

\section{Point spread function}
\begin{figure}
\begin{center}
     \includegraphics[width=.49\textwidth]{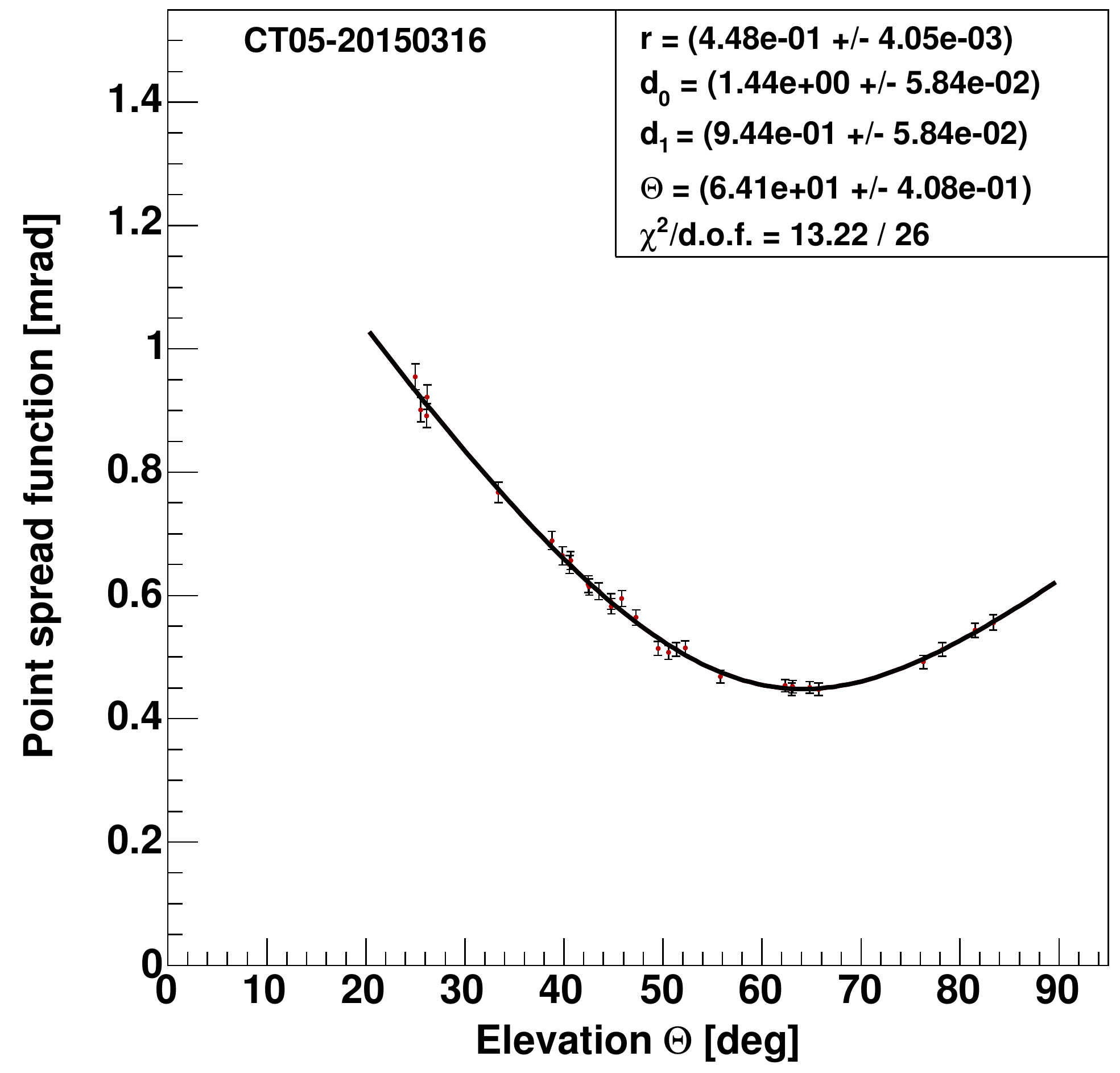}
     \includegraphics[width=.49\textwidth]{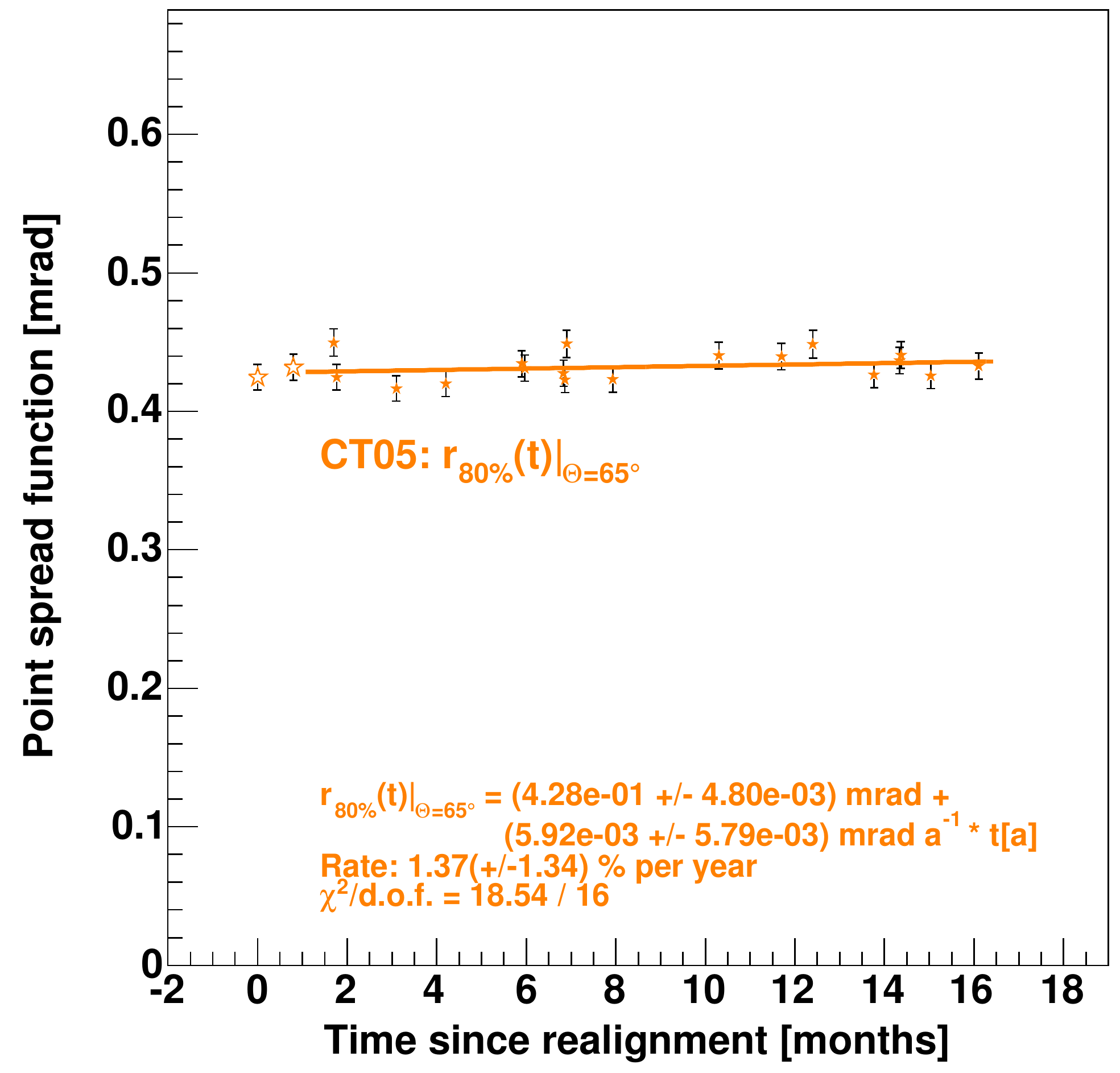}
 \end{center}
    \caption{Optical PSF measurements in one night (left) and since the last realignment(right): The plot on the left hand side is showing individual PSF measurements for different elevation angles of the telescope. The black curve is a fit of Function 5.1 to the data to model the dish deformation. On the right hand side the evaluated Function 5.1 at \SI{65}{\degree} elevation angle since the last realignment is shown. To the datapoint a linear function was fitted to derivate the increase of the PSF over time.}
     \label{fig4}
     \end{figure}

To determine the quality of the alignment, the point spread function (PSF) is monitored regularly while the telescope is not observing. For this measurement the telescope is pointing to a bright star and an image is taken by the CCD in the dish pointing to the closed lid of the Cherenkov camera. To quantify the width of the intensity distribution of the image of the star on the lid of the Cherenkov camera, a radius $r_{80}$ of a circle around the center of gravity of the image containing 80\% of the total intensity is determined. To characterise the PSF over the entire operational range of the telescope, a series of 30 measurements using stars with different altitude and azimuth angles is done \cite{Cornils}. In Figure \ref{fig4} on the left such a series of measurements is shown. An elevation dependent deformation of the dish structure was expect to impact telescope PSF, but in agreement with our expectation no azimuth dependency has been found. In the distribution a clear minimum between \SI{60}{\degree} and \SI{70}{\degree} can be seen, as expected, since the alignment of the mirrors was done in this elevation range which is the typical observation elevation. Moving away from this ideal elevation angle the PSF widens due to deformations in the dish structure and in the connection between the mirrors and the dish. The $r_{80}$ spans between \SI{0.45} and \SI{0.8}{mrad} in the observation elevation range (\SIrange{30}{80}{\degree}). Since the Cherenkov camera pixel diameter is 1.22 mrad, the obtained values are considered acceptable for the physics performance of the telescope.\\
\section{Stability}
Since for the H.E.S.S. experiment an initial alignment is used for an extended period of time without realignment in between, the stability of the alignment over time is important. To determine the degradation of the PSF over an extended period of time, the PSF is measured at least once every two month by taking a full series of 30 single PSF measurements using various stars under different azimuth and altitude angles in one night. After fitting the $r_{80}$ to the 30 CCD images a plot of the $r_{80}$ over the elevation angle is created. To parameterized the dish deformation, a function describing the spot size $r_{80}$ as a function of the elevation angle $\theta$ is applied. The function has the parameters minimal spot size $r$, elevation angle of the minimal spot size $\theta_0$, the dish deformation parameter relative to the minimum for smaller elevation angle $d_0$ and for larger elevation angles relative to the minimum $d_1$.

\begin{align} 
r_{80}(\theta)&=\sqrt{r^2+d_0^2\cdot(\sin\theta-\sin\theta_0)^2+d_1^2\cdot(\cos\theta-\cos\theta_0)^2}
 \end{align}

An example for this fit is shown in Figure \ref{fig4}. The parameters retrieved from this fit are first of all used as an input for the optical ray tracing simulation used in Monte Carlo simulations to evaluate the performance of the instrument and derive expected values to which the measured events are compared. In addition, the function is evaluated at \SI{65}{\degree} elevation angle which corresponds to the nominal elevation angle of the alignment. These PSF values are plotted over the time since the last alignment was finished, as shown in Figure \ref{fig4} on the left. To describe the evolution of the size of the PSF as a function of time $r_{80}(t)$, a linear function with the initial size of the PSF $r$ and a slope $a$ describe the change of the size per year.

\begin{align} 
r_{80}(t)&=r\cdot a
 \end{align}

The last full alignment of the telescope was finished on the 3rd of March in 2014 and the first full measurement was done in the night from the 4th to the 5th of March in 2014. The achieved PSF right after the alignment was \SI[separate-uncertainty = true]{0.428(5)}{mrad} at \SI{65}{\degree} elevation angle and so far an increase in size of the $r_{80}$ of \SI[separate-uncertainty = true]{1.37(134)}{\percent\per yr} is measured. The PSF is stable over time which means that there is no immediate need for a realignment on foreseeable time scales.\\
In January 2015 a full test of the mechanical system was performed mainly to avoid damage to the actuators due to a long period of downtime. After all actuators have been moved in both direction away from the alignment position and back the measured PSF shows no significant change. 

\section{Conclusions}
The achieved PSF is within expectations and meets the instrument requirements. For a telescope's elevation angle of \SI{60}{\degree} to \SI{70}{\degree}, where the alignment was done, the size of the PSF is smaller than the pixels size of the PMT camera of \SI{1.22}{mrad}. For typical gamma-ray observations between \SI{30}{\degree} and \SI{80}{\degree} elevation angle the diameter of the light spot, defined as the diameter of a circle containing \SI{80}{\percent} of the light, spans between \SI{0.9}{mrad} and \SI{1.6}{mrad}. So far, we do not measure a significant increase of the PSF after more than \SI{1.5}{yr} of operations with the current alignment. The time needed for a full alignment is in the expected range and the software, mechanical components, and electronics fulfill all expectations. 

\section*{Acknowledgement}
The support of the Namibian authorities and of the University of Namibia in facilitating the construction and operation of H.E.S.S. is gratefully acknowledged, as is the support by the German Ministry for Education and Research (BMBF), the Max Planck Society, the German Research Foundation (DFG), the French Ministry for Research, the CNRS-IN2P3 and the Astroparticle Interdisciplinary Programme of the CNRS, the U.K. Science and Technology Facilities Council (STFC), the IPNP of the Charles University, the Czech Science Foundation, the Polish Ministry of Science and Higher Education, the South African Department of Science and Technology and National Research Foundation, and by the University of Namibia. We appreciate the excellent work of the technical support staff in Berlin, Durham, Hamburg, Heidelberg, Palaiseau, Paris, Saclay, and in Namibia in the construction and operation of the equipment.

\end{document}